\documentstyle[12pt]{article}
\begin{document}
\title{Fractional quantization of Hall resistance as a consequence of mesoscopic feedback}
\author{Artur Sowa\\
109 Snowcrest Trail,
Durham, NC 27707 \\
www.mesoscopia.com }
\date{}
\maketitle
\begin{abstract}

A nonlinear single-particle model is introduced, which captures the characteristic of systems in the quantum Hall regime. The model entails the magnetic Shr\"odinger equation with spatially variable magnetic flux density. The distribution of flux is prescribed via the postulates of the mesoscopic mechanics (MeM) introduced in my previous articles [cf. J. Phys. Chem. Solids, 65 (2004), 1507-1515; J. Geom. Phys., Vol. 55/1 (2005), 1-18]. The model is found to imply exact integer and fractional quantization of the Hall conductance. In fact, Hall resistance is found to be $R_H = \frac{h}{2e^2}\frac{M}{N}$ at the filling factor value $N/M$.  The assumed geometry of the Hall plate is rectangular. Special properties of the magnetic Shr\"odinger equation with the mesoscopic feedback loop allow us to demonstrate quantization of Hall resistance as a direct consequence of charge and flux quantization.  I believe results presented here shed light at the overall status of the MeM in quantum physics, confirming its validity.

\end{abstract}

\section{The conceptual foundation of the model} 

In this section I will present a derivation of the model from the principles of quantum mechanics and the postulates of the MeM. Naturally, a faithful mathematical model is of value independently of its theoretical justification. The impatient reader, who is interested in the modeling aspects only, may jump directly to the next section. However, I would like to emphasize that the underlying concepts have much broader consequences than this single result. To begin with, consider a system consisting of a two-dimensional electron gas confined to a surface $S$, say, a rectangular plate.  Suppose now that a magnetic field is applied transversally to the surface. Let the corresponding vector potential be given as
\[
\nabla_A =i\nabla +\frac{e}{\hbar} A .
\] 
We focus on those electronic properties of the system, which can be captured by the single-particle Schr\"odinger equation
 \begin{equation}
 i\hbar\dot\Psi = H_A\Psi,
 \label{regSchrod}
 \end{equation}
where the magnetic Hamiltonian $H_A= -\frac{\hbar ^2}{2m^*}\nabla _A^*\nabla_A$. 
The state vector $\Psi$ is assumed to fully (or sufficiently) characterize the electronic state. In particular, the total free charge is given as 
\begin{equation}
Q= e \int\limits_{S}|\Psi |^2.
\label{Charge}
\end{equation}
(The adopted convention is that the unit of $|\Psi |^2$ is $[m^{-2}]$, which in turn affects the definition of current density, etc.) In equilibrium, $\Psi=e^{-iEt/\hbar}\Psi _0$  where $\Psi _0$ is time-independent and satisfies the eigenvalue problem
 \begin{equation}
 H_A\Psi _0 = E\Psi _0.
 \label{eigSchrod}
 \end{equation}
At this point we will use an assumption which is extrinsic to the classical quantum mechanics. Namely, the mesoscopic mechanics postulates (cf. \cite{sowa6}, Second Postulate of the MeM) that the planar distribution of the magnetic flux is probabilistically described in terms of a certain transform $K$ applied to the electronic state $\Psi $. Namely, the magnetic flux density is given as 
\begin{equation}
B(x)= \Phi|K\Psi |^2(x),\quad x\in S,
\label{fill-general}
\end{equation}
where $\Phi$ is the total magnetic flux through the surface, i.e. the number of quanta of the magnetic field.  The magnetic flux density function $x\rightarrow B(x)$ affects the electronic states as it modifies the Hamiltonian $H_A$, thus closing the mesoscopic feedback loop. $H_A$ is modified via the vector potential, which in a simply-connected domain satisfies $\nabla \times A = B(x)$. The particular distribution prescribed in (\ref{fill-general}) is the effect of an interaction of the magnetic field with the electronic structure. The interaction is described via an operator evolution equation called the Mesoscopic Schr\"{o}dinger equation
\begin{equation}
\label{MesoSchr}
i\hbar\dot{K} = -KH_A  -B^2 (K^*)^{-1} ,
\end{equation}
which yields the transform $K$. The constant $B$ is different than zero when the field is switched on, but its exact value plays no explicit role in the analysis below.  The unique structure of solutions of equation (\ref{MesoSchr})  has been described in \cite{sowa6}. Throughout this article we focus attention on operator states (transforms) $K$ of a special type, the equilibrium type, which satisfy the eigenvalue problem
\begin{equation}
KH_A  +B^2 (K^*)^{-1} = \nu K.
\label{Euler-Lagrange}
\end{equation} 
 The constant $\nu$ is interpreted as the Fermi energy level.  It is easily seen, cf. \cite{sowa5} or \cite{sowa6}, that equilibrium states assume the form
\begin{equation}
K = U \circ\sum_{E_n < \nu}\frac{B}{(\nu-E_n)^{1/2}}\quad |\psi _n\rangle\langle \psi_n|.
\label{theK}
\end{equation}
The vectors $|\psi _n\rangle$ represent electronic eigenstates of $H_A$, corresponding to eigenvalues $E_n$. State vectors $|\psi _n\rangle$ are mutually orthogonal, which possibility is guarantied by the hermicity property of the Hamiltonian. (We remark that an arbitrary choice is involved if there are multiple eigenvalues.) The non-Abelian phase matrix $U$ is unitary and  we will assume here that in fact $U=I$ (identity). Let us briefly remark that the $U=I$ equilibrium solutions are interpreted within the MeM framework as representing the phase-correlated state, cf. \cite{sowa5} and \cite{sowa6} for details. Furthermore, we note that $K$ may require normalization so as to guarantee that the probabilistic measure in (\ref{fill-general}) has total mass one. This is achieved by adjusting the constant $B$ if necessary. Now, with the $K$ variable all set, recall that $\Psi _0$ is on the list of eigenstates. We assume that $E_0=E<\nu$, which guaranties that $\Psi _0$ can be identified as one of the vectors in formula (\ref{theK}). Of course, if there are multiple eigenvalues this latter circumstance may require a change of basis.  The arrangements we have made guarantee  that the magnetic feedback loop is switched on. Indeed, formula (\ref{fill-general}) yields 
\begin{equation}
B(x)= \Phi|K\Psi |^2(x) = b |\Psi _0|^2,
\label{fill-frac}
\end{equation}
where the constant $b$ is such that 
\[
\Phi= \int\limits_{S} B(x) = \int\limits_{S} b|\Psi _0 |^2 = b Q/e = bN.
\]
Here, $N$ is the number of relevant (current carrying) charge quanta. In addition, let $M$ denote the number of magnetic flux quanta, so that the total flux $\Phi = Mh/(2e)$. Quantization of flux and charge implies that $b$ is a natural fraction expressed in units of the flux quantum, i.e.
\begin{equation}
b = \frac{M}{N}\frac{h}{2e}.
\label{rational-b}
\end{equation}
The fraction $N/M$ is known in the literature as the filling factor, cf. \cite{Chakraborty}. (The filling factor is typically discussed in the context of constant-magnetic-field Hamiltonian and the Landau levels.) Next we note that in a simply connected region, such as the rectangular surface $S$, the vector potential can be represented in a suitable gauge as 
\[
A = \varphi _x dy-\varphi_y dx.
\]
Note that formula (\ref{fill-frac}) is equivalent to
\begin{equation}
\varphi _{xx} +\varphi_{yy} = b|\Psi _0 |^2. 
\label{Poisson}
\end{equation}
In addition, the eigenvalue problem (\ref{eigSchrod}) expressed in terms of $\varphi$ assumes the form
\begin{equation}
 -\hbar ^2(\partial ^2_x +\partial ^2_y)\Psi _0 + 
2ie\hbar (-\varphi _y\Psi_{0,x} +\varphi _x\Psi _{0,y})+e^2(\varphi _x^2+\varphi _y^2)\Psi_0 = 2m^*E\Psi _0 .
\label{eig-explicit}
\end{equation}
Therefore an application of the postulates of the MeM has lead us to a nonlinear system consisting of equations (\ref{Poisson}) and (\ref{eig-explicit}). Nonlinear systems are prominently present in the framework of quantum mechanics and (the classical) Maxwell theory. However, the novelty of the MeM is here manifested in the particular form of the nonlinearity. One might note that in the classical setting a Poisson equation of the from (\ref{Poisson}) would yield the electric potential, yet here the variable $\varphi$ is tied to the vector potential. In the remainder of this article I will argue that this system of equations provides a faithful model for electronics of the QHE and FQHE regimes.

\section{Calculation of the Hall resistance}

As demonstrated above the MeM prompts us to consider the magnetic Schr\"odinger equation (\ref{eig-explicit}) jointly with the Poisson-type equation (\ref{Poisson}).  
 At this point, we will attempt to understand some properties of this nonlinear system of equations.  To begin with, let us impose an Ansatz  
\[
\Psi_0(x,y) = e^{ikx} \chi (y),
\]
where $\chi = \chi (y)$ is a real function, and the unit of $\chi$ is $[1/m]$. 
Let us also simplify the gauge representation of vector potential $A$ by asking that $\varphi =\varphi (y)$. In such a case (\ref{Poisson}) implies
\begin{equation}
\varphi _y = b \int\limits_{0}^y dy' \chi ^2 (y').
\label{phint}
\end{equation}
Also, evaluating the Hamiltonian we obtain
\begin{equation}
\begin{array}{rl}
2m^*e^{-ikx}H_A\Psi_0 = & \hbar ^2k^2\chi -\hbar ^2\chi_{yy}  
+ 2e\hbar k\varphi _y\chi+e^2\varphi _y^2\chi \\
& = -\hbar ^2\chi_{yy} +\left(\hbar k +e\varphi _y\right)^2\chi .
\end{array}
\end{equation}
Thus, system (\ref{Poisson})--(\ref{eig-explicit}) has been reduced to a nonlinear integro-differential eigenvalue problem, namely:
 \begin{equation}
 -\chi '' + \left(k +\frac{e}{\hbar}b \int\limits_{0}^y dy' \chi ^2 (y')\right)^2\chi = \frac{2m^*E}{\hbar ^2} \chi .
\label{eq-chi}
\end{equation}
It is convenient to introduce an auxiliary variable $w = k +\frac{e}{\hbar }b \int\limits_{0}^y dy' \chi ^2 (y')$. Equation (\ref{eq-chi}) becomes a pendulum-type system
\begin{equation}
\label{w-system}
\begin{array}{rl}
w'= & \frac{e}{\hbar } b \chi ^2 \\
\chi '' = & (w^2-\frac{2m^*E}{\hbar ^2}) \chi  .
\end{array}
\end{equation}
In the next section we will discuss solutions of this equation so as to show that there exist physically significant wavefunctions satisfying the system (\ref{Poisson})--(\ref{eig-explicit}). First, however, let us calculate the resulting Hall resistance. Let us recall that the Hall plate $S$ is a rectangle. Assume that the edges of $S$ are parallel to the $x$ and $y$-axes. The $x$-extent of the rectangle plays no explicit role in the analysis, but the $y$-extent figures more explicitly in our calculation. We will refer to the cut-offs localized at fixed $y$-coordinates as the `left edge' and the `right edge'. Now, let us note that the Hall potential, which is the difference of potentials between the two edges of the rectangle $S$, is directly read out of the Hamiltonian. (We will see in the next section that in fact for certain solutions the Hamiltonian exhibits broken symmetry.) More precisely, the Hall potential is directly derived from the reduced equation (\ref{eq-chi}). It amounts to 
\begin{equation}
\label{Hall-potential}
V_H= \frac{\hbar ^2}{2m^*e}(w^2(\mbox{right edge}) -w^2(\mbox{left edge})).
\end{equation}
We shall see that this is the only possible interpretation. Indeed, a straightforward calculation shows that only the longitudinal (i.e. $x$-) component of the resulting current density is nonvanishing. In fact, it amounts to 
\[
j_x=\frac{e}{m^*}\Re \left\{\Psi _n^*\left(i\hbar \nabla +eA \right)\Psi _n\right\} = \frac{e\hbar}{m^*}\left(k +\frac{e}{\hbar }b \int\limits_{0}^y dy' \chi ^2 (y')\right)\chi ^2.
\]
Thus, potential $V_H$ is transversal to the current. It is now useful to represent the current density in the auxiliary variables. Namely,
\begin{equation}
\label{longitudinal-current}
j_x= \frac{e\hbar}{m^*} w \chi ^2 = \frac{\hbar ^2}{m^*} \frac{1}{b}w w' = \frac{\hbar ^2 }{m^*} \frac{1}{2b}(w^2)'.
\end{equation}
Integrating and taking into account (\ref{Hall-potential}), we obtain that the total longitudinal current $I_x$ is given as
 \begin{equation}
\label{longitudinal-total}
I_x= \frac{e}{b}V_H.
\end{equation}
In view of (\ref{rational-b}) we conclude that the Hall resistance $R_H=V_H/I_x$ is given by the following formula:
\begin{equation}
\label{FQHE}
R_H = \frac{h}{2e^2}\frac{M}{N}.
\end{equation}
This is our main result. It would be interesting to understand the physical significance of factor $2$ in the denominator, but I have no further comments  about it at present. 

\section{Existence of the mesoscopic FQHE states}

We need to show that there exists a wavefunction $\Psi _0$, which (in pair with a matching $\varphi$) satisfies the system (\ref{Poisson})--(\ref{eig-explicit}) and is square-integrable in the rectangle $S$. In view of the reductions we have introduced in the previous section it now suffices to show that equation (\ref{eq-chi}) admits a bounded solution. This is what we will in fact observe. Remarkably, we will see that the solution conforms (although not strictly coincides) with the London penetration law as it applies to the magnetic flux density and the current density in superconducting materials, cf. e.g. \cite{Ibach}. Indeed, let us focus on the dynamical system (\ref{w-system}). Differentiating the first equation we observe that
\[
\frac{w''}{w'}=2\frac{\chi '}{\chi}.
\]  
Substituting into the second equation we obtain an autonomous equation for $w$, namely:
\begin{equation}
\label{w-autonomous}
\frac{w'''}{w'}=\frac{1}{2}\left(\frac{w''}{w'}\right)^2+2w^2-2\tilde{E}
\end{equation}
where $\tilde{E}=2m^*E/\hbar ^2$. The first equation in (\ref{w-system}) shows that as long as $\chi$ does not vanish $w$ is a monotonous function. Thus we can assume without loss of generality that $w'$ is a function of $w$, namely
\begin{equation}
w'=f(w).
\label{f-first}
\end{equation}
Equation (\ref{w-autonomous}) is now represented in the form
\begin{equation}
ff_{ww}+\frac{1}{2}f_w^2 = 2(w^2 - \tilde{E}).
\label{second-order}
\end{equation}
A particular solution (of the form we need) is now easy to guess. Indeed, assume 
\[
f(w) = \alpha w^2+\beta w +\gamma.
\]
Substituting this into (\ref{second-order}) we obtain $f=(\sqrt 2 /2)w^2 -\sqrt 2 \tilde{E}$. Next, substituting $f$ into (\ref{f-first}) we derive a particular solution of interest, i.e. 
\[
w = -\sqrt{2\tilde{E}}\frac{1+Ce^{-2\sqrt{\tilde{E}}y}}{1-Ce^{-2\sqrt{\tilde{E}}y}},
\]
where $C$ is a constant such that $Cb>0$. We now calculate via (\ref{w-system}) that the wavefunction factor $\chi$ is given by
\begin{equation}
\label{chi-London} 
\chi (y)= \pm 2\cdot 2^{.25}\sqrt{\frac{\tilde{E}C\hbar}{eb}}\frac{e^{-\sqrt{\tilde{E}}y}}
{1-Ce^{-2\sqrt{\tilde{E}}y}}. 
\end{equation}
This is a bounded function on any right half-axis, and in particular between the two edges of the Hall plate. Recall that the wavefunction is determined as
\[
\Psi = e^{-iEt/\hbar}e^{ikx} \chi (y).
\]
Thus $\Psi$ is square-integrable over the rectangle $S$, which is what we have intended to prove. In addition, note that constants $C$ and $\tilde{E}$ are in fact constrained by condition (\ref{Charge}). Suppose for example that the left edge of the plate is at $y=0$, while the right edge is sufficiently far to the right so as to make the exponential factor $\exp{(-\sqrt{\tilde{E}}y)}$ negligible. The value of $C$ is then easily estimated to yield
\[
C \simeq 1/(1+\frac{ 2\sqrt{2\tilde{E}}L_x}{M\pi}).
\]
Here $L_x$ denotes the length of the plate along the $x$-direction. Let us emphasize the remarkable fact that charge density function $|\Psi |^2$ is localized at the edge of the plate, and its depletion toward the center is essentially indistinguishable quantitatively from the London penetration law. Here, the depth of penetration is inverse proportional to the square root of energy $E$. This localization property carries over to the current density, cf. formula (\ref{longitudinal-current}), as well as the magnetic flux density, cf. (\ref{fill-frac}). 

In this section we have demonstrated that there are physically meaningful solutions of the set problem. As explained in the previous section, the electronic phenomenon they describe is characterized by fractionally quantized Hall resistance.

\section{Closing comments}

It is worthwhile to notice that the model introduced here does not make any direct references to material properties. It is derived from the magnetic Shr\"odinger equation, assuming a special form of charge-to-flux feedback loop. The loop has been postulated as relevant to planar electronic systems in my previous articles on the MeM, cf. \cite{sowa5}, \cite{sowa6}. In this article our discussion is limited to a single special (if important) case derived from a rich structure of the MeM. 
I would also like to mention that I investigated the consequences of a similar mesoscopic feedback loop for the first time in \cite{sowa7}. That paper treats the \emph{strongly correlated} case as captured by the formalism of a theory called the (fully) Nonlinear Maxwell Theory,  cf. \cite{sowa4}, which is in some sense dual to the MeM. The reader may wish to consult the differences and similarities of the two approaches and their consequences, e.g. equation (\ref{w-autonomous}) above has a companion of similar-type exhibited in the other paper. 

Experimental discovery of the quantum Hall effect was announced in \cite{Klitzing}, while the fractional quantum Hall effect was first announced in \cite{Tsui}. Over the years these phenomena have been discussed theoretically from many vantage points, cf. e.g. \cite{Laughlin}; the literature is very abundant. The particular theme of spatially variable magnetic fields has been considered in the context of FQHE by many authors, e.g. \cite{Kalmeyer}, \cite{Falko}, and \cite{Huckestein}. More precisely, those papers examine the effect of randomly fluctuating fields. 
In comparison, here I have examined a postulate that the spatial distribution of the magnetic field depends in a certain way upon charge distribution. I believe there is now compelling evidence for the validity of this approach.

\end{document}